\pdfoutput=1
\documentclass[preprint,prb,amsmath,superscriptaddress,aps,subeqn,showpacs]{revtex4}
\usepackage{graphicx}

\begin{document}

\title{First-Principles Calculation of Thermal Transport in the Metal/Graphene System}

\author{R. Mao}
\affiliation{Department of Electrical and Computer Engineering, North Carolina State University, Raleigh, NC 27695-7911}

\author{B. D. Kong}
\affiliation{Department of Electrical and Computer Engineering,
North Carolina State University, Raleigh, NC 27695-7911}

\author{C. Gong}
\affiliation{Department of Materials Science and Engineering, The University of Texas at Dallas, Richardson, Texas 75080, USA}

\author{S. Xu}
\affiliation{Department of Physics, North Carolina State University, Raleigh, North Carolina 27695-8202, USA}

\author{T. Jayasekera}
\affiliation{Department of Physics, Southern Illinois University, Carbondale, IL, 62901}

\author{K. Cho}
\affiliation{Department of Materials Science and Engineering, The University of Texas at Dallas, Richardson, Texas 75080, USA}
%\affiliation{Department of Physics, The University of Texas at Dallas, Richardson, Texas 75080, USA}

%\author{M. Buongiorno Nardelli}
%\affiliation{Department of Physics and Department of Chemistry, University of North Texas, Denton, TX %76203} \affiliation{CSMD, Oak Ridge National Laboratory, Oak Ridge, TN 37831}

\author{K. W. Kim}\email{kwk@ncsu.edu}
\affiliation{Department of Electrical and Computer Engineering, North Carolina
State University, Raleigh, NC 27695-7911} %\date{Oct. 2012}% %\revised{\today}%

\begin{abstract}

Thermal properties in the metal/graphene (Gr) systems are analyzed
by using an atomistic phonon transport model based on Landauer formalism and
first-principles calculations.  The specific structures under investigation
include chemisorbed Ni(111)/Gr, physisorbed Cu(111)/Gr and Au(111)/Gr,
as well as Pd(111)/Gr with intermediate characteristics.  Calculated results
illustrate a strong dependence of thermal transfer on the details of interfacial
microstructures.  In particular, it is shown that the chemisorbed case provides
a generally smaller interfacial thermal resistance than the physisorbed due to
the stronger bonding.  However, our calculation also indicates that the weakly chemisorbed
interface of Pd/Gr
may be an exception, with the largest thermal resistance among the considered.
Further examination of the electrostatic potential and interatomic force constants reveal
that the mixed bonding force between the Pd and C atoms results in incomplete
hybridization of Pd and graphene orbital states at the junction, leading effectively
to two phonon interfaces and a larger than expected thermal resistance.  Comparison with available experimental data shows good agreement.  The result clearly suggests the feasibility
of phonon engineering for thermal property optimization at the interface.

\end{abstract}


\maketitle

\section{Introduction}

As the critical dimensions of modern electronic devices approach the nanoscale,
the power density of integrated circuits has soared drastically, forcing the
issue of thermal management to the forefront.  Faced with the challenge, graphene has
recently emerged as one of the key candidates for the next-generation low-power
electronics for its superb properties.  Fundamental understanding of thermal transport
in the graphene based structures is crucial from the perspective of both low-dimensional
physics and practical applications of this emerging material system.  Particularly,
the metal contact with graphene is of interest as it provides not only
an essential part of any active device but also a primary path of heat dissipation.
Thermal conduction across the heterogeneous metal/graphene (Gr) interface is characterized
by the interfacial resistance often known as the Kapitza resistance.~\cite{Pollack1969}
The heat current is mainly carried by phonons, while the electronic contribution
is considerably smaller. It was found recently that thermal energy transfer via direct
electron-phonon coupling astride the heterointerface is insignificant (compared to the
phonon-phonon interactions) in metal/dielectric structures including those employing graphene.~\cite{Lyeo2006,Hopkins2012}

As in the case of most layered structures, the properties of the metal/graphene system
are also expected to depend heavily on the bonding chemistry and detailed
structures at the interface.  It was recently
demonstrated that the bonding between graphene and metal atoms can be divided into two
categories$-$chemisorption and physisorption.~\cite{Khomyakov2009} While
chemisorption opens a band gap in graphene due to hybridization between
the graphene $p_z$- and metal $d$-orbitals, the Dirac-cone feature of
graphene is preserved at the physisorbed interfaces.  Subsequent investigations
illustrated that physisorption is observed for Ag, Al, Cu, Cd, Ir, Pt, and Au,
whereas the Ni, Co, Ru, Pd, and Ti interfaces belong to chemisorption.~\cite{Ran2009,Gong2010,Gong2012}
However, their impact on phonon/thermal transport has received
much less attention with only a limited number of studies available in the literature.
A molecular dynamics calculation was conducted for heat transfer between allotropic
carbon nanotubes and Cu substrate.~\cite{Gao2011}  On the side of experiment,
measurement of thermal conductance was reported for the Cu/Gr, Al/Gr, Ti/Gr
and Au/Gr cases with the obtained values ranging $ 0.8 - 2.5 \times 10^{-8}$ Km$^2$/W.~\cite{Schmidt2010}

%The above research have either only focused
%on one particular metal material or resort to classical DMM models for
%phonon transport.~\cite{Swartz1989,Little1959}
In this paper, we attempt to provide a detailed theoretical
account of interfacial thermal resistance in the metal/graphene system. The
sample structures are chosen to reflect the range of typical interfaces
from chemisoprtion to physisorption for a comprehensive analysis.  Since atomistic
details of phonon dynamics are crucial for the accurate outcome, the adopted theoretical
approach utilizes a first-principles analysis based on density functional theory
(DFT) to calculate the interatomic force constant (IFC) matrix.~\cite{Baroni2001,Gonze1997,Alam2011}
Then the desired phonon transmission function and the thermal current are determined
via Green's function techniques~\cite{Lee1981,Datta1997,Nardelli1999,Allen1993} and the Landauer formalism.~\cite{Landauer1970}  The obtained results are compared with experimental
and other theoretical data available in the literature.

The rest of the paper is organized as follows.  First, the structures of
metal/graphene systems under investigation are discussed briefly with the focus
on the interfacial bonding environment. Then, a summary description of the adopted thermal
transport model is provided including the details of numerical implementation.
Finally, discussion/analysis of the calculated interfacial phonon transmission and
Kapitza resistance is presented.

\section{Metal/graphene interfaces}
\begin{center}
\begin{figure}
\includegraphics[width=8cm]{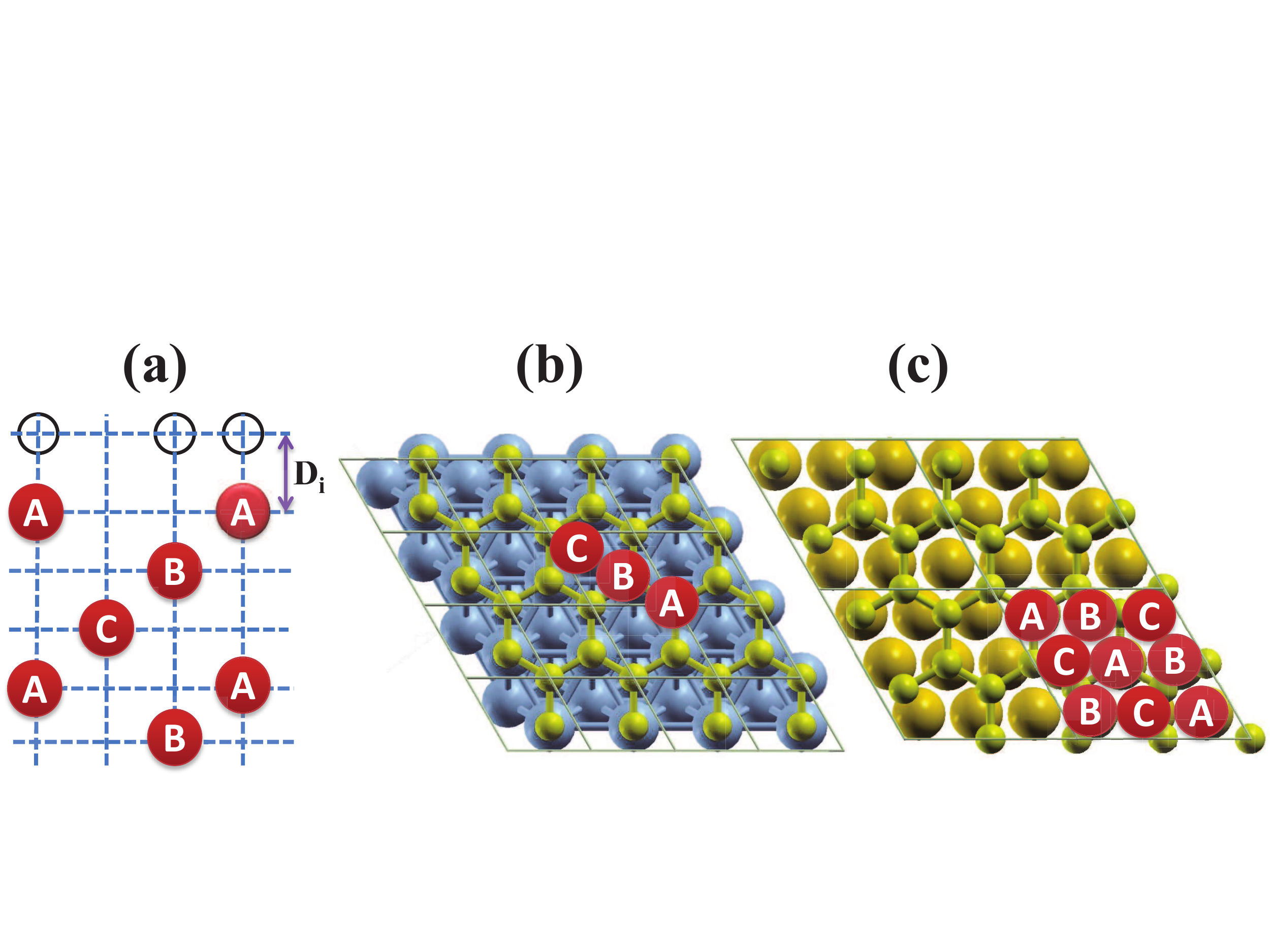} %bb=48 314 750 532,
\caption{(Color online) Schematic representation of (a) stacking order
and (b,c) top view of graphene absorbed on the metal (111) surface.
(b) corresponds to Ni/Gr and Cu/Gr, while (c) is for Pd/Gr
and Au/Gr. $D_{i}$ specifies the interfacial separation between
graphene and the metal.} \label{Schematic}
\end{figure}
\end{center}

\begin{center}
\begin{figure}
\includegraphics[width=8cm]{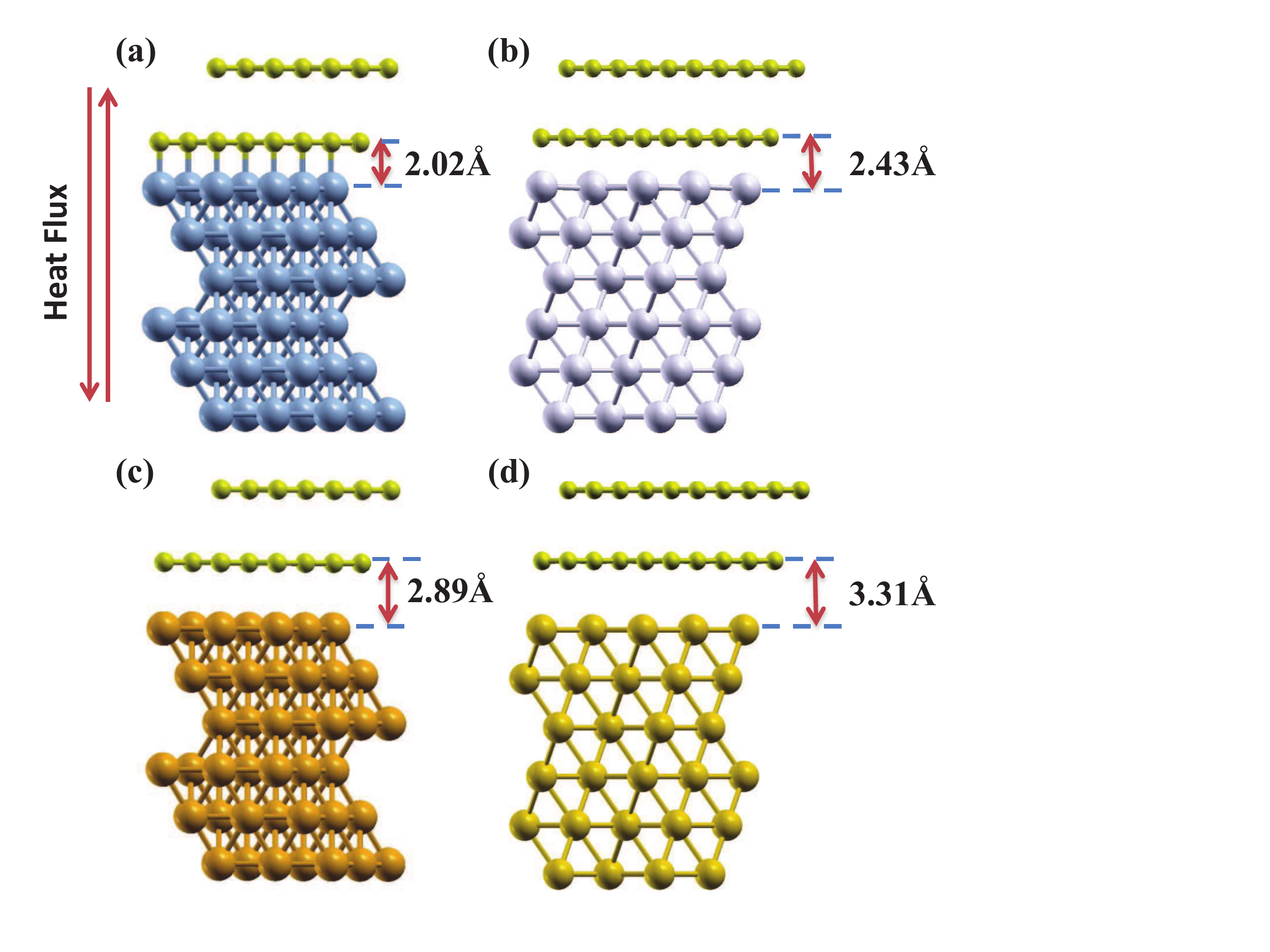} %[bb=16 388 543 684, width=8.5cm]
\caption{(Color online) Side view of the investigated metal/graphene
systems; (a) Ni(111)/Gr, (b) Pd(111)/Gr, (c) Cu(111)/Gr, and (d) Au(111)/Gr.
Two upper layers represent graphene.} \label{Structure}
\end{figure}
\end{center}

The heterogeneous system of interest is graphene on the (111) surface
of a metal (Ni, Pd, Cu, and Au).  If the lattice constant of graphene
is fixed at the optimized value 2.46 \AA, less than 5\% lattice mismatch is
introduced when these metals of face-centered cubic symmetry are made commensurate
with the graphene lattice.  A 1$\times$1 unit cell is formed for Ni/Gr and Cu/Gr,
while a 2$\times$2 construction is necessary for Pd/Gr and Au/Gr to accommodate larger
sizes of metal atoms.
The top view and stacking orders of the four different material combinations
are shown schematically in Fig.~\ref{Schematic}, representing the ideal atomic arrangement
at the metal/graphene contacts.  While useful in illustrating the geometric
construction, these arrangements are insufficient to account for details of the
interfacial microstructures.  Instead, the realistic structures can be obtained through
geometry optimization using first-principles calculation (see, for example, Fig.~\ref{Structure}).  As indicated earlier, the relaxed metal/graphene interfaces are generally divided into two categories, chemisorption and physisorption, depending on the bonding energies, interfacial separations, and orbital hybridizations.  In accord with earlier studies,~\cite{Gong2010} our calculation clearly illustrates that Ni and graphene bond strongly at the interface through hybridization between Ni $d$-orbitals and C $p_z$-orbitals.  Strong coupling also results in a relatively small interfacial separation (2.02 \AA).  On the other hand, Cu and Au are physisorbed on graphene and form a weak van der Waals bonding with a larger interlayer distance (2.89 {\AA}
and 3.31 {\AA}, respectively).~\cite{comm2}  For the Pd(111)/Gr structure, it is observed that the interaction between Pd and C atoms may be smaller than the strong chemical bonding, leading effectively to the mixed character at the interface (i.e., of both chemisorption and physisorption).~\cite{Ran2009} This unique combination may set the Pd/Gr system apart from the other metal/graphene interfaces as elaborated further in Sec.~IV.

\section{Theoretical Model}

\subsection{Thermal transport in the atomistic Green's function formulism}

Phonon transport is originated from the dynamics of the lattice or
lattice vibrations.~\cite{Ashcroft1976} Since the investigation concerns the
interfacial properties, transport in the immediate region astride the interface
can be treated ballistic and a quantum mechanical treatment
in the Landauer framework~\cite{Landauer1970} is adequate to capture the
essential features.  Accordingly, we consider a three parted system where the central
interface region (i.e., the region of interest) is connected to the thermal
reservoirs on the left and the right with two semi-infinite leads (labeled L and R),
often known as the lead-conductor-lead configuration.~\cite{Nardelli1999,Zhang2007}
Accounting for only the phonon contribution, the thermal current density is then
given as:
\begin{equation}\label{thermalcurrent}
J(T_\mathrm{L},T_\mathrm{R})=\int_0^{+\infty}\frac{d\omega}{2 \pi} \hbar \omega
{\mathcal T}_{ph}(\omega) \left[ n(T_\mathrm{L},\omega) - n(T_\mathrm{R},\omega)\right],
\end{equation}
where $n (T_\mathrm{L,R},\omega)$ is the equilibrium phonon distribution in
the left or right thermal reservoir (at temperature $T_\mathrm{L,R}$) and ${\mathcal T}_{ph}(\omega)$
denotes the phonon transmission function.
In this expression, ${\mathcal T}_{ph} (\omega)$
is directly related to the lattice dynamics of the given structure containing
all of the relevant details.  Its computation can be achieved in a manner analogous
to that of electron transmission coefficients across nanoscale heterostructures by using
such approaches as the atomistic Green's function method.~\cite{Nardelli1999,Zhang2007,Calzolari}

%The equilibrium atomistic
%arrangement of the device, especially the interfaces, thus have a
%vital role in determining how phonons transport through the system.
%By modifying the interfacial chemistry or geometry, the phonon
%transport properties can be engineered. $T(\omega)$ can be
%calculated using .  The
%technical details of the calculation are equivalent to
%that used for the electron transmission coefficient calculation
%across nanoscale interfaces.

The first step of calculating the thermal current [i.e., ${\mathcal T}_{ph}(\omega)$]
is to obtain the IFCs defined as:
\begin{equation}\label{ifceq}
K_{i,j,\alpha,\beta} (\textbf{R})=\frac{\partial ^2 E
(\textbf{R})}{\partial u_{i,\alpha}
\partial u_{j,\beta}}~~ \text{for}~ i\ne j \,,
\end{equation}
where $E$ is the total energy of the system, $\textbf{R}$ is a Bravais lattice vector,
and $u_{i,\alpha}$ is
the displacement of the $i^{th}$ atom (of the unit cell) in the $\alpha $ direction
with respect to the equilibrium position.  The $i=j$ cases can be determined in terms of
those in Eq.~(\ref{ifceq}) via the acoustic sum rule.~\cite{Ashcroft1976}
The IFCs in turn define the harmonic matrix
\begin{equation}
\tilde{K}=\{\tilde{K}_{i,j,\alpha,\beta}\}=K_{i,j,\alpha,\beta}/\sqrt{M_i M_j}
\end{equation}
and the dynamical matrix
\begin{equation}\label{FourierTransform}
D_{i,j,\alpha,\beta}(\textbf{q})=\frac{1} {\sqrt{M_i M_j}}\sum_{\textbf{R}}
K_{i,j,\alpha,\beta}(\textbf{R})e^{-i \textbf{q} \cdot \textbf{R}} \,,
\end{equation}
where $M_i$ is the mass of the $i^{th}$ atom.
%The phonon frequencies $\omega$
%are given as solutions of the eigen value problem:
%\begin{equation}\label{latticedynamics}
%\text{det} \left|  D_{i,j,\alpha,\beta} (\textbf{q}) - \omega^2 (\textbf{q})
%\right| =0 .
%\end{equation}
%The calculation of IFCs is done by utilizing the Density Functional
%Perturbation Theory (DFPT) to obtain the dynamical matrices in the
%$\textit{k}$ space and then Fourier transforming to the real space.
In the considered lead-conductor-lead system, this harmonic matrix can be written
formally as~\cite{Zhang2007,Calzolari}
\begin{align} \label{harmonicM}
\tilde{K}= \begin {pmatrix}
\tilde{K}_\mathrm{L} & \tilde{K}_\mathrm{LC} & 0 \\
\tilde{K}_\mathrm{LC}^\dag & \tilde{K}_\mathrm{C} & \tilde{K}_\mathrm{RC} \\
0 & \tilde{K}_\mathrm{RC}^\dag & \tilde{K}_\mathrm{R} \end {pmatrix}
\end{align}
and, finally, the dynamical equation as
\begin{equation} \label{dynamicalE}
\text{det} \left | \tilde{K}-\omega ^2 I \right |=0 \,.
\end{equation}
Here, $\tilde{K}_{p}$ is the matrix in each of the three regions ($ p=$ L,C,R), while
$\tilde{K}_{p\mathrm{C}}$ represents the phonon coupling at the interfaces between the
central region and the two leads ($p =$ L,R).  This expression in Eq.~(\ref{dynamicalE})
is clearly analogous to the governing equation of the electronic system,
with $\omega^2 \leftrightarrow
\epsilon$ and $\tilde{K} \leftrightarrow H $ [see, for example, Eq.~(1) of
Ref.~\onlinecite{Nardelli1999}].  Accordingly, calculation of phonon transmission
can directly follow the Green function techniques developed initially for
electronic transport as mentioned earlier.

Utilizing the identified parallelism, the phonon transmission function is  written
straightforwardly as:
\begin{equation}\label{transmission}
{\mathcal T}(\omega)= \mathrm{Tr} \left( \Gamma_\mathrm{L} G^r_\mathrm{C}
\Gamma_\mathrm{R} G^a_\mathrm{C} \right ),
\end{equation}
where $G^{r,a}_\mathrm{C}$ are the retarded (r) and advanced (a) Green's
functions of the central region and $\Gamma_{\mathrm{L,R}}$ correspond to the coupling
with the left and right leads, respectively.  Further, the Green's
function and the coupling functions can be obtained from~\cite{Calzolari}
\begin{equation}
G^r_\mathrm{C}=\left[(\omega + i 0)^2 - \tilde{K}_\mathrm{C} - \Sigma^r_\mathrm{L}
-\Sigma^r_\mathrm{R}\right]^{-1} \,,
\end{equation}
\begin{equation}\label{couplingmatrix}
\Gamma_p=i[\Sigma^r_p-\Sigma^a_p] \,,
\end{equation}
where the self-energy terms
\begin{equation}\label{selfenergies}
\Sigma^r_{p}= \tilde{K}_{p\mathrm{C}}^\dagger G_p^r \tilde{K}_{p\mathrm{C}}
\end{equation}
are evaluated with the help of the Green's function $G^r_p$ in the corresponding
leads ($p=$ L,R).  The transfer matrix method offers a very efficient approach to
calculate $G^r_p$ in the semi-infinite lead by treating it as a stack of principal
layers with only nearest-layer interactions.~\cite{Lee1981,Sancho1985,Nardelli1999}
The advanced terms (with superscript $a$) are given as Hermitian conjugates.  Additional
details of underlying theoretical formulation can be found in Ref.~\onlinecite{Calzolari}.

Once the phonon transmission function $\mathcal{T}(\omega)$ is obtained,
thermal and thermoelectric properties can be evaluated.  A particularly
relevant quantity is the thermal resistance $R_{th}$ of the system that is given
as the inverse of the thermal conductance:
\begin{align}\label{thermalconduct}
R_{th} (T)= \kappa_{ph}^{-1} (T) = \left[ \int_0^{+\infty}\frac{d\omega}{2 \pi} \hbar \omega
{\mathcal T}_{ph}(\omega) \frac{\partial n(T,\omega)}{\partial T} \right]^{-1} .
\end{align}
While the model described above is for the case of two leads (or reservoirs),
extension to a multi-terminal system is trivial as it has been demonstrated in electronic transport.~\cite{Jayasekera2007}

\subsection{Numerical implementation}

The calculations are performed in the DFT framework, as it is implemented in the
QUANTUM-ESPRESSO package,~\cite{Giannozzi2009} with ultrasoft pseudopotentials in
the local density approximation.~\cite{comment} A minimum of 35 Ry is used for
the energy cut-off in the plane wave expansion along with the charge truncation of
350 Ry. In addition, the Methfessel-Paxton first-order spreading is employed
with the smearing width of 0.01 eV. The momentum space is sampled on a Monkhorst-Pack
mesh in the first Brillouin zone.  In the construction of interfacial structure,
two layers of graphene and three to five layers of metal (depending on the material) are
considered in the calculation.  The unit cell is set to the lattice constant of
graphene and the geometry optimization is performed to find the energy minimum structure.
The Hellman-Feynman force is taken into account in each iterative solution.
Figure~\ref{Structure} shows the resulting interface structures of the considered material
combinations. They serve as the central region in the previously mentioned lead-conductor-lead configuration.  Two leads consisting of respective bulk materials (i.e., bulk metal and graphene/graphite) connect seamlessly to the interface region and are modeled separately.
No appreciable mismatch (i.e., resistance) exists between the leads and the conductor. Simulation
results show that the k-space grids of 6$\times$6$\times$2 and 6$\times$6$\times$6 are sufficient
for the interface region and the bulk leads, respectively, with good convergence in the relevant characteristics.

With the optimized realistic interface structures, the next step is to evaluate vibrational
properties including IFCs. While analytical force constant models have been used widely in
such materials as carbon or silicon based systems,~\cite{Al-Jishi1982,Saito1998,ZhangMingo2007}
they are not directly applicable in the current investigation due to the lack of information on the required bonding details between carbon and metal atoms.  Instead, we calculate lattice dynamics,
also within the DFT formalism for accuracy, by utilizing a perturbative treatment known as density functional
perturbation theory (DFPT).~\cite{Baroni2001}  The IFCs are then obtained by Fourier analyzing
the set of dynamical matrices generated from the first-principles calculation under the harmonic approximation [see Eq.~(\ref{FourierTransform})].
With the IFCs of bulk leads, we can build $\tilde{K}_\mathrm{p}$, ($ p=$ L,R)
that enter Eq.~(\ref{harmonicM}); While the on-site matrix $\tilde{K}_{\mathrm{C}}$
and coupling matrices $\tilde{K}_{L\mathrm{C}}$ and $\tilde{K}_{R\mathrm{C}}$
can be constructed from the IFCs of the central
part. Then the phonon transmission can be evaluated using Eq.~(\ref{transmission})
with the Green's function calculated from the transfer matrix technique.  Finally, the
thermal resistance of the structure and the Kapitza resistance at the metal/graphene
junction can be evaluated by Eq.~(\ref{thermalconduct}).  Since this quantity of interest is defined in the near-equilibrium condition (i.e., an infinitesimally small temperature gradient across the structure; see Eq.~(\ref{thermalconduct})], an equilibrium treatment is adequate with no need for an iterative solution.

%The interfacial thermal resistance, following the scenario described
%in Ref.~\onlinecite{Laikhtman1994}, can be obtained as follows:

\section{Results and Discussion}

\begin{center}
\begin{figure}
\includegraphics[width=8cm]{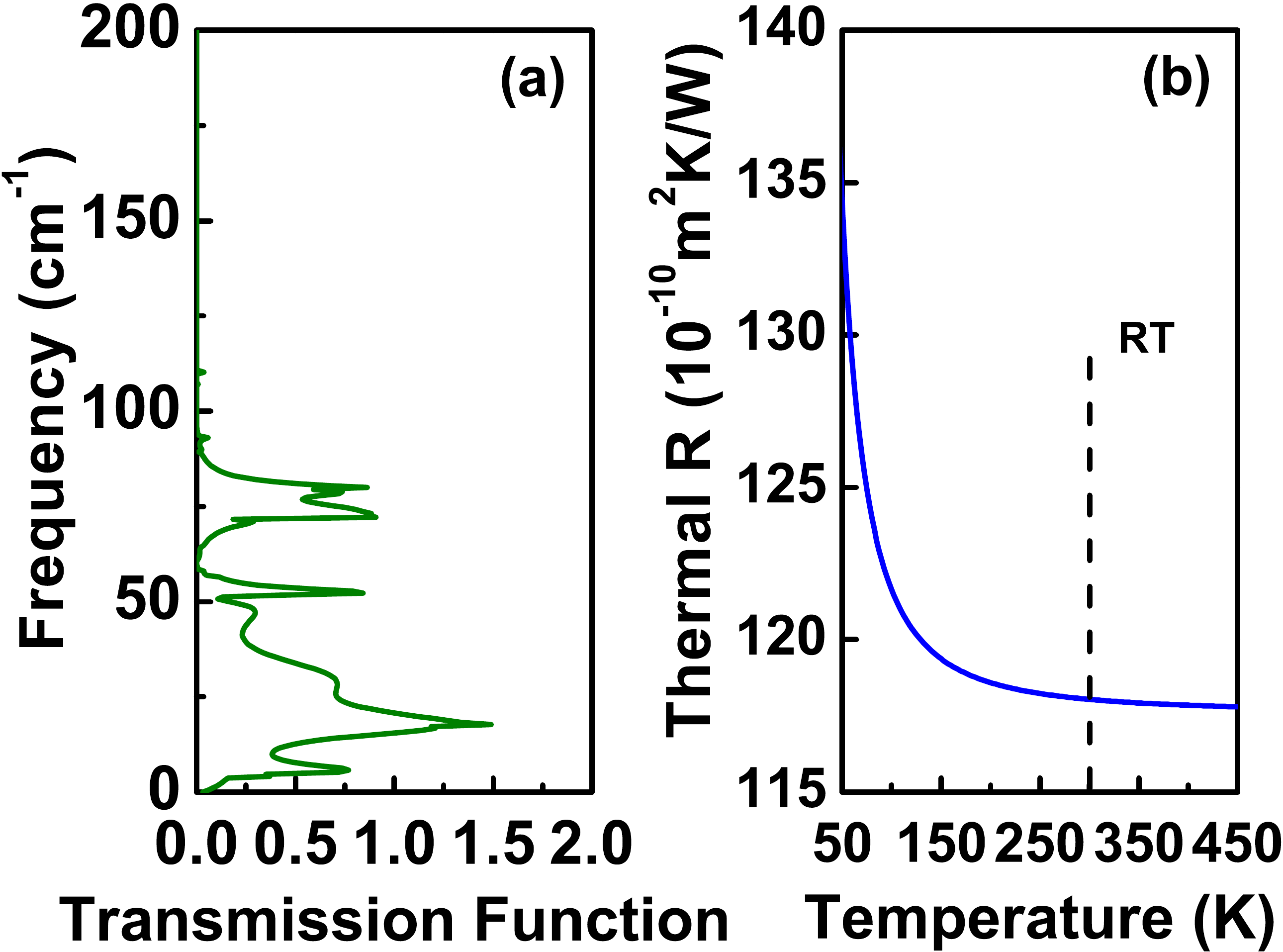} %bb=0 13 240 174,
\caption{(Color online) (a) Phonon transmission function vs.\ frequency and (b) interfacial
thermal resistance vs.\ temperature for Cu/Gr. The vertical dashed line
marks the resistance at room temperature (300 K).}
\label{TransResistCuG}
\end{figure}
\end{center}

Of the material combinations under investigation, graphene on the Cu (111) surface is examined
first due, partly, to its widespread use.  As indicated by the obtained phonon transmission function in Fig.~\ref{TransResistCuG}(a), only the low lying acoustic branches (below 100 cm$^{-1}$) play the dominant role in phonon transport at the interface.  The impact of optical branches is orders of magnitude smaller.  The resulting thermal resistance is plotted in Fig.~\ref{TransResistCuG}(b).  Since the total resistance of the structure [i.e., Eq.~(\ref{thermalconduct})] contains the contribution from the leads as well, the intrinsic thermal resistance at the junction is deduced by subtracting this  portion in a manner analogous to electrical transport.~\cite{Datta1997,Laikhtman1994}  At T=300 K, the estimated interfacial thermal resistance of the Cu/Gr structure is $1.18 \times 10^{-8}$ Km$^2$/W.   The interfacial resistance exhibits the 1/T dependence in the low temperature region (50$-$150 K), while staying almost invariant between 150 K and 450 K.

%(denoted $R_{in}$ in the figure)
%that is actually very close to the recent MD result of approx. $1.2 \times 10^{-8}$ %Km$^2$/W.~\cite{Chang2012}
%Our calculated result is also close to experimentally observed thermal
%resistance of exfoliated Gr on SiO$_2$ ($56-120 \times 10^{-10}$ Km$^2$/W
%at room temperature).~\cite{Chen2009} This may be attributed to the fact that only
%weak bondings are formed at Cu/Gr interface due to the physisorption.

\begin{center}
\begin{figure}
\includegraphics[width=8cm]{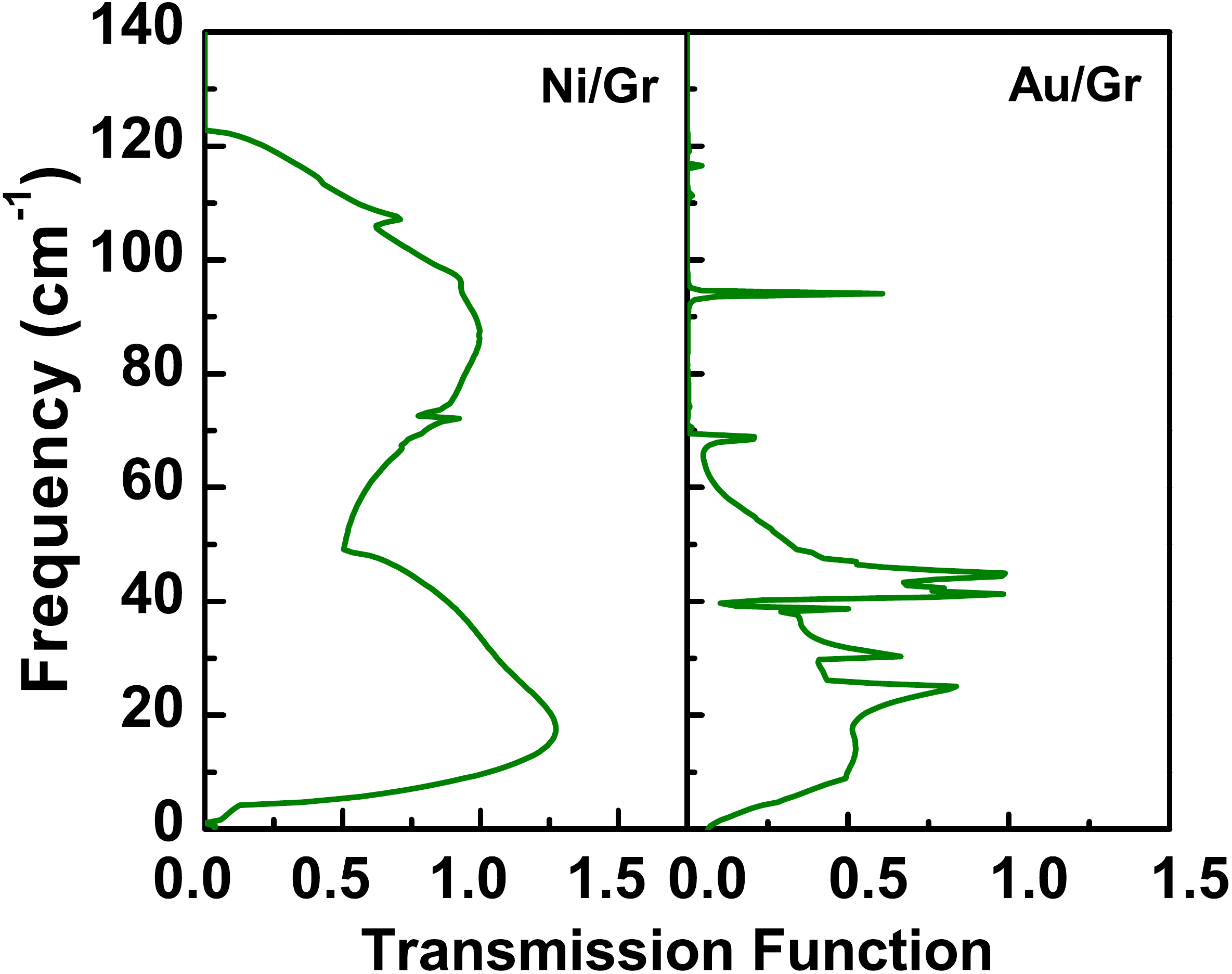}
\caption{(Color online) Phonon transmission function vs.\ frequency for Ni/Gr and Au/Gr.}
\label{TransNiG_AuG}
\end{figure}
\end{center}

\begin{center}
\begin{figure}
\includegraphics[width=8cm]{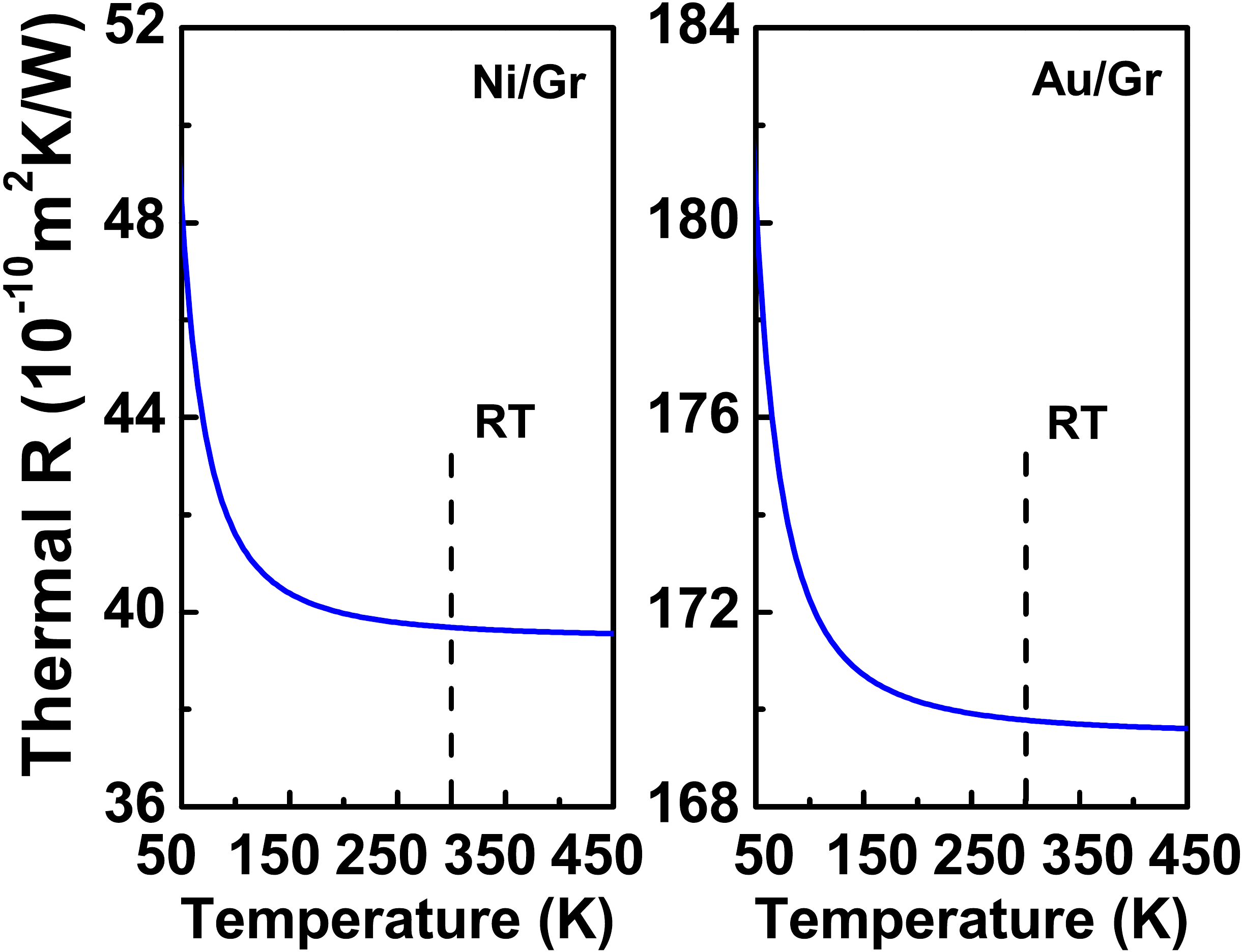}
\caption{(Color online) Interfacial
thermal resistances vs.\ temperature for Ni/Gr and Au/Gr. The vertical dashed lines
mark the resistances at room temperature (300 K).}
\label{RinNiG_AuG}
\end{figure}
\end{center}

When the calculation is extended to chemisorbed Ni and physisorbed Au, the respective results appear to be similar in many aspects.  However, one interesting point to note is that the Ni/Gr interface shows the phonon transmission coefficient whose frequency dependence is much broader with fewer resonant features (see Fig.~\ref{TransNiG_AuG}).  This is substantially different from those of the physisorbed metal/graphene interfaces (both Au and Cu).  The mixed nature of phonon dynamics at the chemisorbed interface (thus, a smaller mismatch) is thought to be the main origin of enhanced phonon transmission and eventually a smaller interfacial resistance.  The estimated value for Ni/Gr is about $3.9 \times 10^{-9}$ Km$^2$/W at 300K, whereas it is more than four times larger for Au/Gr ($1.7 \times 10^{-8}$ Km$^2$/W) as indicated in Fig.~\ref{RinNiG_AuG}.  The obtained result for physisorbed Au/Gr shows good agreement with the measurement data available in the literature ($ \sim 2-3 \times 10^{-8}$ Km$^2$/W).~\cite{Schmidt2010,Hopkins2012}  In contrast, the conventional diffuse mismatch model shows large discrepancies.  Even with the added sophistication such as anisotropy in graphitic materials and multiple heat transfer mechanisms, it substantially overestimates the thermal resistance for Au/Gr ($\sim 6.8 \times 10^{-8}$ Km$^2$/W).~\cite{Duda2010}  A similar calculation yields $\sim 4 \times 10^{-8}$ Km$^2$/W at the Cu/Gr interface.~\cite{Chang2012}

\begin{center}
\begin{figure}
\includegraphics[width=8cm]{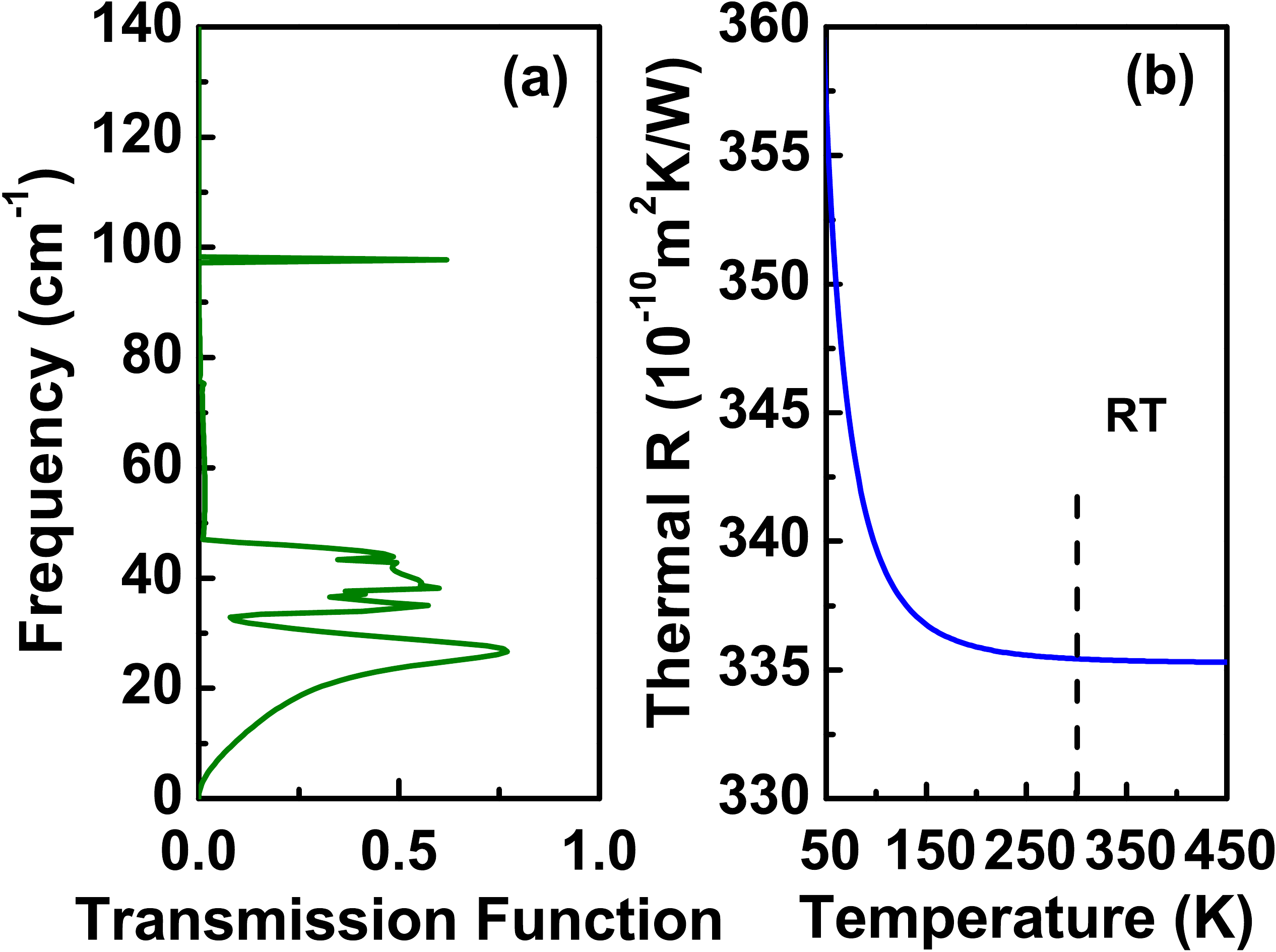}
\caption{(Color online) (a) Phonon transmission function vs.\ frequency and (b) interfacial
thermal resistance vs.\ temperature for Pd/Gr.  The vertical dashed line
marks the resistance at room temperature (300 K).} \label{TransResistPdG}
\end{figure}
\end{center}

Comparison between different graphene/metal systems thus far clearly indicates that the chemisorbed interface is generally more favorable than the physisorbed in term of thermal transport.  The presence of strong bonding and the smaller interlayer separation (e.g., 2.02 {\AA} of Ni/Gr vs. 3.31 {\AA} of Au/Gr) all support this conclusion that is also in accord with a recent experimental study.~\cite{Schmidt2010}
One potential exception may be graphene on the Pd (111) surface.  Since their bonding characteristics supposedly show both chemisorbed and physisorbed nature as mentioned earlier, it is reasonable to anticipate that the interfacial thermal resistance would fall in between as well.   However, the calculation suggests that Pd may not follow the trend and actually have the largest resistance of those considered (a value of $3.35 \times 10^{-8}$ Km$^2$/W at room temperature; see Fig.~\ref{TransResistPdG}).  Most notably, phonon transmission is greatly suppressed between approx. 40 cm$^{-1}$ and 100 cm$^{-1}$ even when compared to Cu/Gr and Au/Gr.  This result is counterintuitive, particularly when the interlayer distance (which tends to indicate the interaction strength) behaves as expected; namely, between the values of chemisorbed and physisorbed structures as indicated in Fig.~\ref{Structure}.

\begin{center}
\begin{figure}
\includegraphics[width=8.5cm]{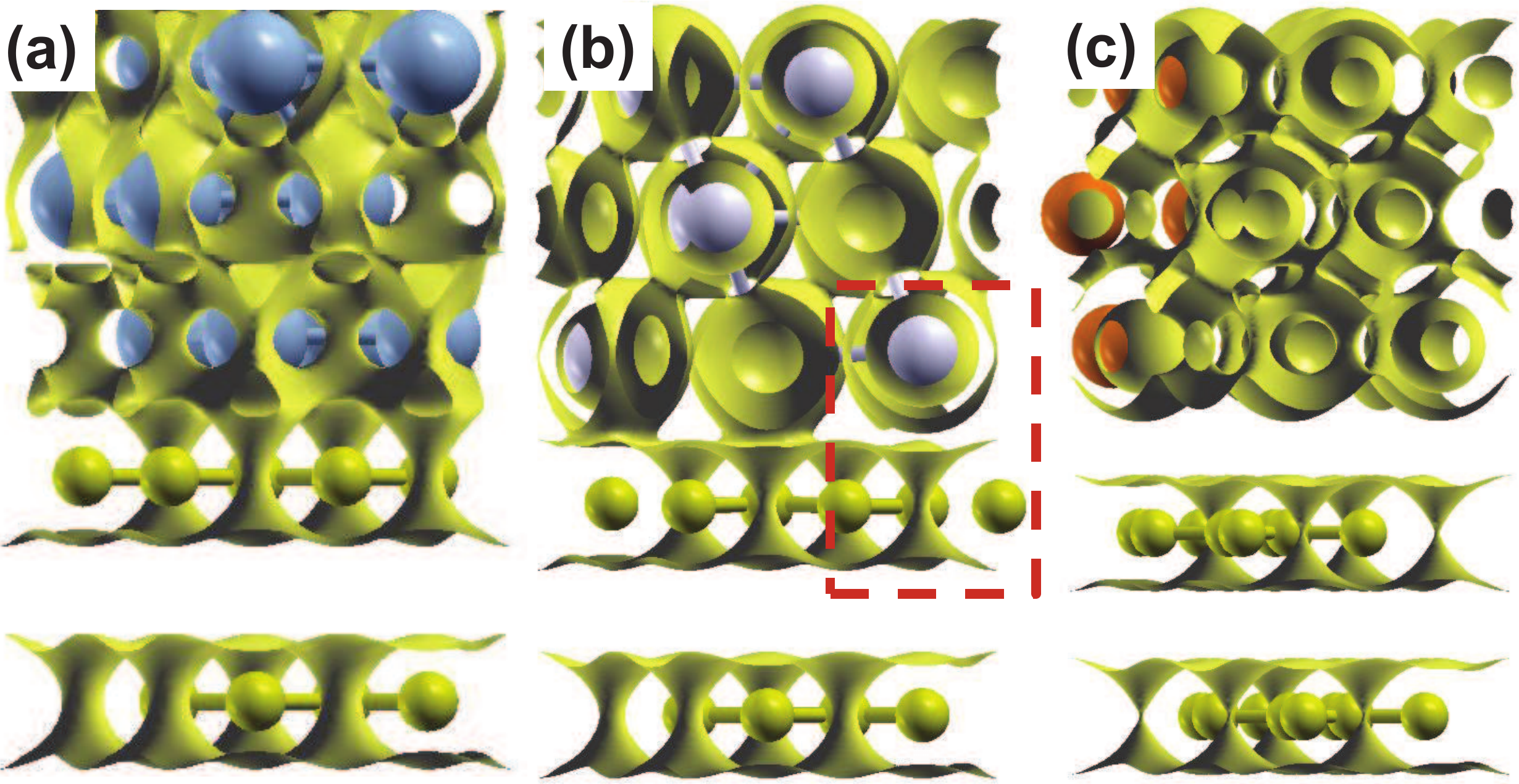}
\caption{(Color online) Electrostatic potential isosurface of the metal/graphene system for
(a) Ni/Gr, (b) Pd/Gr, and (c) Au/Gr. Top three layers are for the metal, while bottom two correspond to graphene.  In (b), partly detached bonding in the boxed region indicates incomplete mixing at the Pd/Gr interface.  The large (colored) balls denote the metallic atoms.}
\label{ElectronicPotent}
\end{figure}
\end{center}

The identified peculiarity is examined further by analyzing detailed microstructures of the corresponding metal/graphene interfaces.  The electrostatic potential isosurface plotted in Fig.~\ref{ElectronicPotent} clearly illustrates that strong hybridization between the  $p_z$ and $d$ orbitals at the chemisorbed Ni/Gr interface "glues" Ni and graphene together, making them essentially one unit.   As such, propagating phonons transmit through the Ni/Gr interface with relative ease.  On the other hand, the weak interaction between Au and C atoms in the Au/Gr system forms a barrier at the interface, which strongly scatters phonon transmission.  For Pd/Gr, it is revealed that the mixed bonding force between Pd and C atoms is indeed smaller than the strong chemical bonding (namely, chemisorption) as evident from the partly detached bond in the boxed region.  However, the interaction still alters the orbital states of first layer graphene, making them sufficiently distinct from those of second layer graphene.  Consequently, phonons potentially face two interfaces for transmission instead of one.   A force constant analysis between atomic layers can clarify the latter point with numerical certainty.

\begin{center}
\begin{figure}
\includegraphics[width=8.5cm]{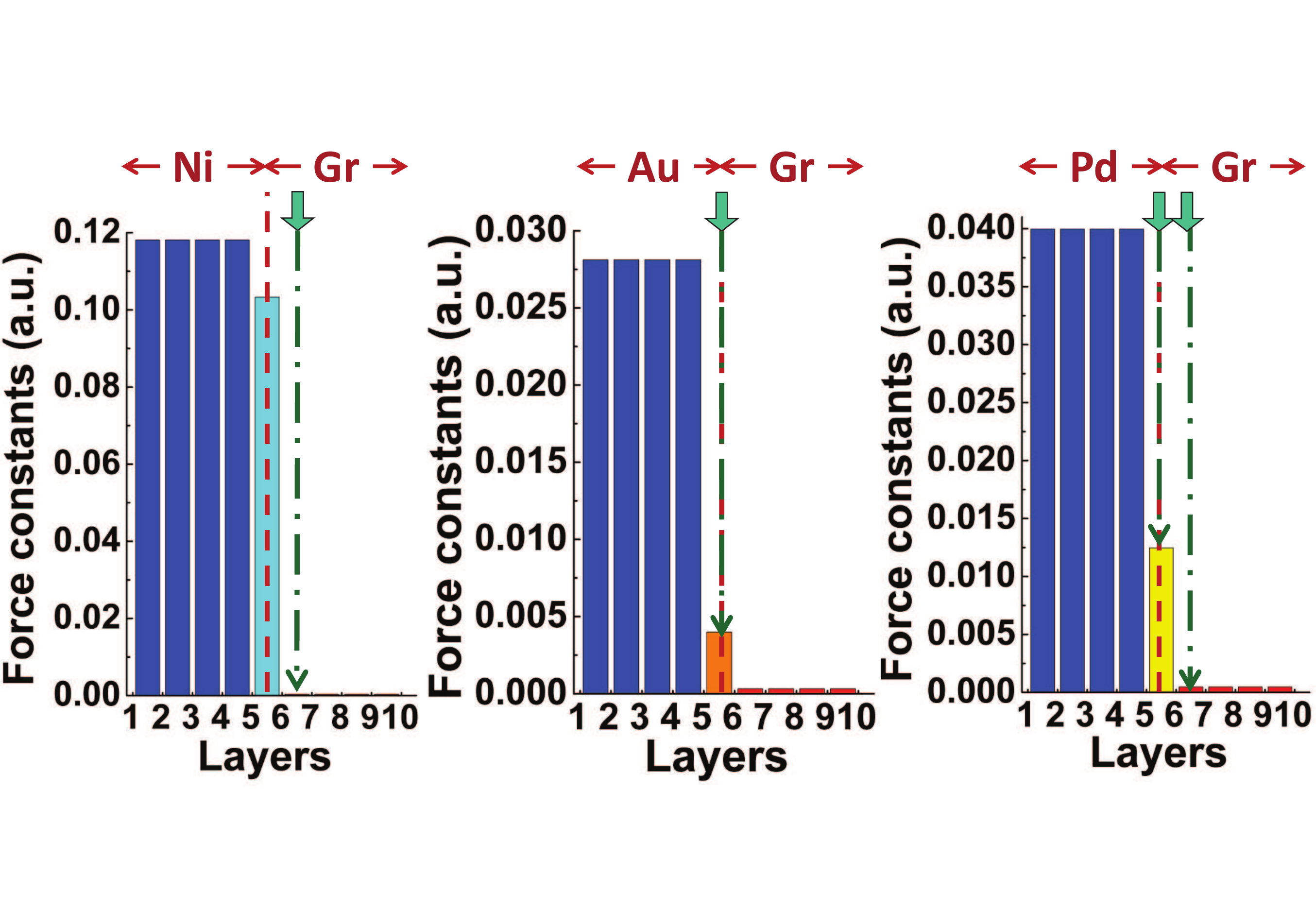} %bb=12 9 254 851,  new bb=3 202 718 484,
\caption{(Color online) Interlayer force constants for Ni/Gr, Au/Gr and Pd/Gr.  The height of each bar represents the interaction strength between two layers, where a.u. stands for atomic Rydberg units.  In all three plots, the metallic layers are up to layer 5 (i.e., 1$-$5) and graphene starts from layer 6 (i.e., 6$-$10). The dashed lines indicate the physical interface between metal and graphene, whereas the dash-dotted lines (as well as the block arrows) symbolize the barriers that transmitting phonons experience.}
\label{Forceconstant}
\end{figure}
\end{center}

Figure~\ref{Forceconstant} illustrates the interlayer force constants deduced from the DFPT calculation.  The height of each bar symbolizes the interaction strength between two neighboring layers.  For example, the first bar on the left denotes the interaction between layers 1 and 2; the next bars are for layers 2 and 3, and so on.  In all three plots, the metallic layers are up to layer 5 and graphene starts from layer 6; accordingly, the physical interface of two heterogeneous materials is located between layers 5 and 6.
On the other hand, the real interface or barrier which impedes phonon transport is characterized by the abrupt change of force constant.
In Ni/Gr, it is illustrating to note that the force constant between the Ni and graphene layers right at the interface (the bar between layers 5 and 6) shows only a slight difference with those between Ni layers on the left (i.e., 0.103 a.u. vs.\ 0.118 a.u., where a.u.\ stands for atomic Rydberg units).  Instead, transmitting phonons experience the major barrier at the interface between layers 6 and 7, where the force constant changes the most drastically.  Due to strong hybridization discussed earlier, the first layer of graphene is absorbed by Ni atoms, leaving it practically decoupled from the second graphene layer.  Nevertheless this is a Gr/Gr interface and it is reasonable to expect a relatively smaller resistance.~\cite{comm3}
For physisorbed Au, the characteristics are as expected. Namely, we get a clear separation of gold bonding and graphene bonding at the interface between layers 5 and 6 (0.028 a.u. vs.\ 0.004 a.u.).  Since gold atoms are much heavier than carbon, acoustic phonon frequencies experience a large mismatch $-$ hence, a large thermal resistance.
When it comes to Pd/Gr, the incomplete mixing of the first graphene layer indeed results in two-step changes in the force constant between layers 5 and 6 as well as between 6 and 7 (i.e., from 0.04 a.u.\ to 0.0125 a.u.\ then to $8.31 \times 10^{-4}$ a.u.), leading effectively to two phonon interfaces as mentioned earlier, and a larger than expected interfacial thermal resistance.
One cautionary point is that the calculation outcome could experience modifications if the graphene film is just one monolayer thick.  Then, the thermal resistance values can be substantially smaller than the presented, particularly when significant mixing is involved.  Our current theoretical formalism is not equipped to address isolated systems, for which the ideal leads are difficult to construct.

Additional insight into interfacial phonon transport may be gained by comparing the results of metal/graphene structures with those involving dielectric substrates.  As summarized in
Table~\ref{Table},~\cite{Mao2012,Chen2009}  a couple of likenesses can be readily noted.  That is, the calculated interfacial thermal resistance of Ni/Gr is close to the corresponding value of BN/Gr ($3.9 \times 10^{-9}$ vs.\ $5.4 \times 10^{-9}$ Km$^2$/W), while Pd/Gr and SiC/Gr are very much alike ($3.35 \times 10^{-8}$ vs.\ $3.61 \times 10^{-8}$ Km$^2$/W).  Since it is the Gr/Gr interface in Ni/Gr that determines the thermal resistance, the chemisorbed case can be understood roughly analogous to the epitaxial Gr/BN structure, also consisting of two materials of similar two-dimensional (2D) crystal type.  In the case of Pd/Gr vs.\ SiC/Gr, both harbor effectively more than one interface/barrier for phonon transmission.  As it is well known, the studied structure of epitaxial graphene grown on the SiC (0001) surface contains an additional carbon buffer layer at the interface that does not have the characteristic sp$^2$ bonding.  Hence, the observed large interfacial thermal resistance in SiC/Gr is consistent with the discussion on Pd/Gr given earlier (i.e., two phonon barriers in series).  Finally, the physisorbed structures with a weak van der Waals bonding between two dissimilar materials (e.g., 3D metal vs.\ 2D graphene) may correspond to exfoliated graphene placed on a non-2D crystal substrate such as SiO$_2$.  While comparable theoretical estimate based on a first principles calculation is not available for SiO$_2$/Gr, the recent experimental data~\cite{Chen2009} of $0.56 - 1.2 \times 10^{-8} $ Km$^2$/W show good match with the values of Cu/Gr and Au/Gr obtained in the current investigation.  These analyses highlight how significantly the atomic bonding details can influence the interfacial thermal properties, leading potentially to phonon engineering for active heat management.~\cite{Hopkins2012,Mao2012}  It also illustrates the inadequacy of various theoretical treatments including the molecular dynamics approach~\cite{Chang2012} that cannot properly account for the required level of physics {\it a priori}.

\section{Summary}

Thermal transport in the metal/graphene heterostructures is investigated by using a first principles
method and the Green's function approach within the Landauer formalism.  The obtained interfacial thermal resistances are $3.9 \times 10^{-9}$ Km$^2$/W, $1.18 \times 10^{-8}$ Km$^2$/W, and $1.70 \times 10^{-8}$ Km$^2$/W at room temperature for the Ni/Gr, Cu/Gr and Au/Gr structures, respectively, indicating generally more effective thermal transfer at the chemisorbed surface owing to the smaller interlayer separation and stronger bonding with graphene.  However, calculations also illustrate that a weakly chemisorbed case such as Pd/Gr could actually lead to an interface even more resistive than that encountered at the physisorbed surface.  Detailed examination of electrostatic potential and force constants  identifies the formation of an intermediate layer (a consequence of incomplete mixing) and the resulting multiple phonon interfaces/barriers as the potential origin of the observed deviation.  Comparison with the corresponding calculations in the graphene/substrate systems reveals strong correlation between seeming differences in material combination, further emphasizing the role of atomic-level {\it ab initio} analysis.  The obtained theoretical results show good agreement with experimental data available in the literature.

\begin{acknowledgments}
This work was supported, in part, by SRC/NRI SWAN and NSF NERC ASSIST (EEC-1160483).
\end{acknowledgments}

\clearpage
\begin{table}
\caption{Thermal properties at the relevant graphene/metal and graphene/substrate interfaces.  The entries for SiO$_2$/Gr are from experiments, whereas the other are from the first-principles calculation.}
\setlength{\tabcolsep}{12pt}
\centering
    \begin{tabular}{c c c c} %{c c c c{5cm}}
        \hline\hline
        ~      & Interface             & Interfacial             & Thermal resistance
\vspace{-8pt}	\\
		   & environment		   & separation				 &  ($10^{-10}$ Km$^2$/W)              \\ \hline %[0.5ex]%
        Ni/Gr   & Chemisorption         & 2.02 {\AA}                & 39                           	\\
        Cu/Gr   & Physisorption         & 2.89 {\AA}                & 118                          	\\
        Au/Gr   & Physisorption         & 3.31 {\AA}                & 170                          	\\
        Pd/Gr   & Mixed                 & 2.43 {\AA}                & 335                          	\\
        BN/Gr$^a$ & Flat              & 3.43 {\AA}                & 54                           	\\
        SiC/Gr$^a$  & Rough (buffer layer)  & 3.89 {\AA}            & 361                          	\\
        SiO$_2$/Gr$^b$ & Rough          & 4.2 {\AA}                 & 56$-$120                      \\
        %[1ex]%
        \hline\hline
$^a$Ref.~\onlinecite{Mao2012} & &
        \vspace{-8pt}\\
$^b$Ref.~\onlinecite{Chen2009}& & \\
\end{tabular}

\label{Table}
\end{table}

\clearpage

\clearpage

\clearpage

\clearpage

\clearpage

\clearpage

\clearpage

\clearpage


\begin{references}
\bibitem{comm2}  Van der Waals interactions are partly included in our calculation.  Specifically, the molecular interaction originated from the static dipole formation is intrinsically considered.  On the other hand, dynamic phenomena such as the instantaneous dipole formation are currently beyond the scope of the DFT formalism.
\bibitem{comment}
While the relativistic pseudopotentials were also considered for the potential impact of spin-orbit coupling in the case of Au/Gr, the calculation results indicate generally negligible differences.  Accordingly, this effect (i.e., spin-orbit coupling) is ignored in the rest of the study.
\end{references}
\end{document}